\documentclass[12pt]{iopart}

\usepackage{iopams}
\usepackage{makeidx}
\usepackage{epsf}
\usepackage{graphics}
\usepackage{graphicx}
\usepackage{amsfonts}
\usepackage{subfigure}
\usepackage{color}

 \expandafter\let\csname equation*\endcsname\relax 
 \expandafter\let\csname endequation*\endcsname\relax 
 \usepackage{amsmath}

\usepackage[update,prepend]{epstopdf}

\begin{document}

\title{Superfluid flow past an obstacle in annular Bose--Einstein condensates}

\author{M~Syafwan$^1$, P~Kevrekidis$^2$, A~Paris-Mandoki$^3$, I~Lesanovsky$^3$, P~Kr\"uger$^3$, L~Hackerm\"uller$^3$ and H~Susanto$^4$}

\address{$^1$Department of Mathematics, Andalas University, Limau Manis, Padang, 25163, Indonesia}
\address{$^2$Department of Mathematics and Statistics, University of Massachusetts, Amherst, MA 01003-9305}
\address{$^3$School of Physics and Astronomy, University of Nottingham, Nottingham, NG7~2RD, United Kingdom}
\address{$^4$Department of Mathematical Sciences, University of Essex, Colchester, CO4~3SQ, United Kingdom}
\eads{\mailto{hsusanto@essex.ac.uk}, \mailto{lucia.hackermuller@nottingham.ac.uk}}

\begin{abstract}
We investigate the flow of a one-dimensional nonlinear Schr\"{o}dinger 
model with periodic boundary conditions past an obstacle, motivated by recent experiments with Bose--Einstein condensates in ring traps. Above certain rotation velocities, localized solutions with a nontrivial phase profile appear. In striking difference from the infinite domain, in this case there are many critical velocities. At each critical velocity, the steady flow solutions disappear 
in a saddle-center bifurcation. 
These interconnected branches of the bifurcation diagram lead
to additions of circulation quanta to the phase of the associated
solution. This, in turn, relates to the manifestation of persistent
current in numerous recent experimental and theoretical works, the connections
to which we touch upon. 
The complex dynamics of the identified waveforms and the 
instability of unstable solution branches
are demonstrated. 
\end{abstract}


\maketitle

\section{Introduction}


Persistent flow is a remarkable property of macroscopic quantum systems. Bose--Einstein condensates (BECs) in a ring geometry \cite{saue01,gupt05,arno06,hend09,ryu07,and14}
have been shown recently to support circulating superfluid flow \cite{ryu07,rama11,moul12}. The ring trap can have a highly tunable radius and controllable transverse oscillation frequency \cite{heat08,sher11}, which makes such a system ideal for the creation of, e.g., a multiply connected BEC \cite{ryu07,moul12} as well as for applications in interferometry \cite{gust97}.

A characteristic feature associated with superfluidity is 
the existence of a critical velocity above which its breakdown 
leads to the creation of excitations. In experiments
with BECs, evidence for a critical velocity was obtained by moving an obstacle, i.e.\ a tightly focused laser beam, through a BEC \cite{rama99,onof00}. This setting
has been demonstrated to be prototypical for dark soliton formation in 1D \cite{Hakim,rad} (see \cite{enge07} for experiments, although the latter were
only quasi-one-dimensional), and for vortex formation in 2D
\cite{adams1}, which can be thought of as a type of nonlinear Cherenkov radiation. In the case of obstacles in a supersonic flow of the BEC, the formed Cherenkov
cone \cite{caru06,glad07,amo09} transforms into a spatial shock wave consisting of a chain of dark solitons \cite{el06}. The appearance of such radiation in photonic
crystals \cite{science} is yet another illustration of the importance of the fundamental study of critical velocity. The formation of vortex dipoles
in a similar setting was also directly observed experimentally in the work
of~\cite{anderson2010}. The case of a heavy impurity and the associated 
drag force were studied in~\cite{astra}.

In a homogeneous weakly interacting Bose gas the critical velocity is the same as the speed of sound, as per the associated Landau criterion~\cite{pavloff}.
Moving inhomogeneities can alter this critical value. For a ring geometry it has been shown in~\cite{dube12} that the instability of the superfluid is caused by outer and
inner edge surface modes, in a similar fashion as in an infinite cylindrically symmetric tube with transverse harmonic confinement \cite{fedi01,angl01}. The different
mechanism is due to the presence of a centrifugal force arising from the 
nature of the rotation. The effect of potential barriers in BECs confined in a ring trap has been
studied experimentally and theoretically \cite{rama11,piaz09,wright13,eckel14,muno15,yaki15}. The weak link due to the barrier, which affects the current around the loop, has a promising
application as a closed-loop atomic circuit (atomtronics), e.g.\ as analogs of superconducting quantum interference devices (SQUIDs) \cite{ryu13,jendr14}. 
The current-phase relation of a BEC flowing through a weak link
was explored for a repulsive square barrier in~\cite{piazza_extra}.
The existence of a critical velocity
above which superfluid flow stops in the ring 
is connected to Cherenkov radiation through the excitations of vortex-antivortex pairs
\cite{rama11,piaz09}, in analogy to the rectilinear case \cite{anderson2010}. Nevertheless, the relevant instability
remains somewhat inconclusive (in connection to corresponding 
experiments~\cite{rama11}) with different mechanisms proposed to account
for discrepancies between theory and observation including thermal
fluctuations~\cite{mathey12} and imaging system resolution~\cite{piaz13}.

In the present study we consider a BEC confined in an effectively one-
dimensional annular trap with a moving potential barrier, 
which is equivalent to a stationary barrier
and a moving condensate as realized in \cite{rama11, wright13}. Using the mean-field, i.e.\ Gross-Pitaevskii
(GP) approximation in the \textit{infinite} domain, it was shown that the 
critical velocity $v_{c}$ corresponds to a saddle-center bifurcation of two branches of solutions
\cite{Hakim}, a stable (center) and an unstable (saddle) one. Using a 1D approximation (for a narrow ring geometry) 
in an effectively periodic
domain, we reveal in an analytical and corroborate in a numerical
fashion the existence of a sequence of saddle-center bifurcations 
and associated critical velocities.
These, in turn, correspond to different topological charges that are all connected within the same bifurcation diagram. 
We present numerical simulations as well 
as analytical calculations,
where it is shown that the critical points can be obtained from solving two 
coupled nonlinear equations. {We also observe the presence of a critical strength of the inhomogeneity (or length of the 
domain) above (respectively, below) which there is no critical velocity, i.e.\ the inhomogeneity can move with any velocity while preserving the superfluidity.} This occurs when the ring circumference becomes shorter than the healing 
length (a setting that
may thus be less relevant from a physical perspective) 
or when the obstacle is strong enough. 
Our examination reveals a series of unstable branches in the relevant
dynamics; we explore the dynamical evolution of the solutions associated
with these branches by means of direct numerical simulations.


The results presented here are intimately connected with
recent experimental and theoretical observations. One of the early attempts to identify
topological winding (and unwinding) in atomic BECs resulted in the 
seminal findings of~\cite{Kanamoto08,Kanamoto09a,Kanamoto09b}. In these
works, rather than a defect rotating inside a BEC, a setting
where a rotation was imposed on the entire quasi-1D ring BEC 
was examined. This has similarities but also substantial
differences from our setup. A similarity is that the system is
analytically tractable; in fact, it is a genuinely homogeneous
system (1D in the co-rotating frame) where the effective 1D
GP equation associated with the dynamics (including
the rotational term) can be solved analytically by means of
elliptic function cnoidal wave solutions which account for the phase slip events also identified here.
On the other hand, there are nontrivial differences from 
that case. In particular, in our setting (and in recent experiments such 
as \cite{rama11,wright13,eckel14,ryu13,jendr14,eckel14b},
the phase slips do not arise
in a ``distributed'' manner, associated with these periodic
cnoidal solutions, but rather in a localized manner being
co-located with the defect. Hence, the analytical considerations
presented herein are expected to be more closely connected
to recent experiments. Among the latter, a few~\cite{ryu13,jendr14}
have been more directly related to the case with a pair
of weak links or Josephson junctions, aiming at least in part
to potential superconducting quantum interference
(SQUID) related applications, while here we will
focus solely on the realm of a single weak link. Arguably,
the recent experimental settings most clearly related to our own
work are those of~\cite{wright13,eckel14}. The former
one measures experimentally the emergence of the phase
slips and uses a qualitative model based on the Bohr-Sommerfeld
quantization condition and an approximate current-phase
relation at the weak link to theoretically trace a structure similar to
the one that we analytically identify in the present work [cf.
their Fig. 4 and our bifurcation diagrams of Figs. 2 and 3 below]. 
The latter work of~\cite{eckel14}, in fact, explicitly
identifies the hysteretic dynamics that has been
proposed to be a key characteristic of this system in the above
figures. However, it also recognizes the disparity of the experimental
observations from the Gross-Pitaevskii findings (a feature
that retraces discussions of earlier work mentioned 
above~\cite{mathey12,piaz13}). 
The analytical tractability of our
findings in this system (in a sense, adapting
to it the spirit of calculations performed earlier in the
homogeneous rotated system of~\cite{Kanamoto08,Kanamoto09a,Kanamoto09b})
may offer further insight in the relevant comparison. 

As a final
step in this theme of  comparisons, 
we would like to mention recent work, which has explored
the case of a rotating weak link as a function not of the potential/domain
parameters considered here (such as the barrier strength or the
domain length), but rather as a function of the interaction strength~\cite{Cominotti14}.
This elaborate task requires different approaches in the weakly interacting
limit (treated by means of a Gross-Pitaevskii equation) and
in the strongly interacting limit (treated by means of a
Luttinger liquid approach and in the case of a Tonks gas
by a Bose-Fermi mapping to the case of non-interacting
fermions). Intermediate regimes were treated by density-matrix
renormalization group computations which, in fact, revealed
an unexpected optimality in the observed persistent currents
at some intermediate interaction strengths between the
above limits.

\section{Theoretical Setup}

The 3D Gross-Pitaevskii equation is given by \cite{bao13}
\begin{equation}
i \hbar \partial_t \psi =- \frac{\hbar^2}{2m}\Delta\psi +g_{3D}n|\psi|^2\psi+V(\mathbf{x})\psi +U(\mathbf{x},t)\psi,
\label{NLS0}
\end{equation}
with $\mathbf{x}=(x,y,z)\in\mathbb{R}^3$, $\psi(\mathbf{x},t)$ is the mean-field wave function, $m$ is the atomic mass, $\Delta=\partial_{xx}+\partial_{yy}+\partial_{zz}
$ is the Laplacian, $\mu$ is the chemical potential, 
$g_{3D}=\frac{4\pi\hbar^2a_s}{m}$ the atomic interaction strength, which is proportional to the atomic scattering length $a_s$, $n$ is the number of atoms, $V$ is the external trapping potential, 
and $U$ is a short range potential representing the moving obstacle with an angular velocity $\omega$.  Here, the wave function is scaled by the integral 
\[
\int_{\mathbb{R}^3} |\psi|^2\,d\mathbf{x}=1.
\]
BECs on a ring, with radius $R$, can be described by the GP equation (\ref{NLS0}) with a trapping potential that can be written in cylindrical coordinates $(r, z, \theta)$ as 
\[
V(r,z)= \frac{1}{2}m[\omega_r^2(r-R)^2+\omega_z^2z^2]
\]
where $\omega_r,\omega_z\gg1$, such that the dynamics of the BECs would be confined at $r=R$ and $z=0$, 
which is the minimum of the confining potential $V(r,z)$. 
Typical parameters used in experiments (\cite{wright13}) with $^{23}$Na are $\omega_r /2\pi = 110\,$Hz, $\omega_z /2\pi = 550\,$Hz,$R=20\,\mu$m, $n \sim 10^5$ and the Planck constant $\hbar = 1.05 \times 10^{-34}\,$Js.

 In this case, $\Delta=\partial_{rr}+\frac1r\partial_r+\frac1{r^2}\partial_{\theta\theta}+\partial_{zz}$ and the moving potential $U$ can be treated as a $\delta$ potential of strength $\alpha$, i.e., $U(r,\theta,t)=\alpha \delta\left(r-L/\pi\right)\delta(\theta-\omega t).$ 
Assuming a concentration around the minimum $(r=R,\, z=0)$ and using $\omega_{z}, \omega_{r} \gg 1$, one can then write $\psi(\mathbf{x},t)=\psi_1(r)\psi_2(z)\psi_3(\theta,t)$, where $\psi_1(r)$ and $\psi_2(z)$ are the ground states satisfying the equations
\begin{align*}
&\left[-\frac{\hbar}{2m}\left(\partial_{rr}+\frac1r\partial_r\right)+\frac{1}{2}m[\omega_r^2(r-R)^2]\right]\psi_1(r)=\kappa_1\psi_1(r),\\
&\left[-\frac{\hbar}{2m}\partial_{zz}+\frac{m}{2}\omega_z^2z^2\right]\psi_2(z)=\kappa_2\psi_2(z),
\end{align*}
under the scaling $\int_{\mathbb{R}^+}r|\psi_1|^2\,dr=\int_\mathbb{R}|\psi_2|^2\,dz=1$. Substituting $\psi(\mathbf{x},t)=\psi_1(r)\psi_2(z)\psi_3(\theta,t)$ into \eqref{NLS0} one obtains approximately
\begin{align}
i \hbar \partial_t \psi_3(\theta)&=-\frac{\hbar^2}{2m}\frac{1}{R^2}\partial_{\theta\theta} \psi_3(\theta) + g_{3D} n|\psi_1(R)|^2|\psi_2(0)|^2|\psi_3(\theta)|^2\psi_3(\theta)+(\mu_1+\mu_2)\psi_3(\theta) \notag \\ 
&+U(R,\theta,t)\Psi_3(\theta).
\label{NLS01}
\end{align}
Using the scaling $\tilde{t}=t/t_0$ and $\tilde{\psi}=\sqrt{\tilde{g}}\psi_3$,
with
\begin{equation*}
t_0=\frac{\hbar}{2m}, \,\tilde{g}=(g_{3D} n |\psi_1(R)|^2|\psi_2(0)|^2 2m)/(\hbar^2),
\end{equation*}
and setting $\tilde{\kappa}=(\kappa_1 + \kappa_2)t_0/\hbar$, $\tilde{\omega}=\omega t_0$, and $\tilde{\alpha}=\alpha/(\hbar t_0)$ accordingly, Eq.~(\ref{NLS01}) can be reduced into the scaled equation (after dropping all the tildes)
\begin{equation}
i\psi_t=-\frac{1}{R^2}\psi_{\theta \theta} +\kappa \psi + |\psi|^2\psi +\alpha\delta(\theta-\omega t)\psi,
\label{NLS}
\end{equation}
where the periodic boundary conditions along the azimuthal direction are
\begin{equation*}
\psi(-\pi,t)=\psi(\pi,t),\qquad \psi_\theta(-\pi,t) = \psi_\theta(\pi,t).
\label{b1}
\end{equation*}


A static BEC with a moving potential is equivalent to the flow of a nonlinear Schr\"odinger (NLS) fluid past an immobile obstacle. 
Writing $R=L/\pi$, taking $R\theta\rightarrow x$, and considering the travelling frame
(i.e., $x \to x-vt$), Eq.~(\ref{NLS}) for the effectively
1D problem can be written as
\begin{equation}
i\psi_t=iv\psi_x-\psi_{xx} +\kappa \psi + |\psi|^2\psi +\alpha\delta(x)\psi,\,-L\leq x<L.
\label{travNLS}
\end{equation}
In order to study the existence of persistent superflow, we search for a steady state solution $\psi(x,t)=\exp(-i\mu_* t) u(x)$ of the GP equation (\ref{travNLS}), where $\mu_*$ is the chemical potential, which leads to 
\begin{equation}
iv u_x-u_{xx} +\kappa u + |u|^2 u +\alpha\delta(x) u=\mu_* u,
\end{equation}
or equivalently
\begin{equation}
iv u_x- u_{xx} - \mu u + |u|^2u +\alpha\delta(x)u=0
\label{travNLS0}
\end{equation}
with $\mu=\mu_*-\kappa$. 

Denoting $N_L$ as the norm of the solution $u(x)$, i.e.\ $N_L=\int_{-\pi}^{\pi}|u|^2 dx=\tilde{g}$, Eq.\ (\ref{travNLS0}) is solved simultaneously with the scaling equation, without loss of generality, $N_L=2L$, 
using a Newton-Raphson method by discretizing the Laplacian with a central finite difference method~\cite{seyd10}.

We then examine the (linear) stability of a solution $u(x)$ for which we introduce the linearization ansatz 
$\psi(x,t)=\exp(-i\mu_* t)(u(x)+\epsilon(r(x)\mathrm{e}^{i\omega t}+s^{\ast}(x)\mathrm{e}^{-i
\omega^{\ast}t}))$, where $\epsilon$ is a formal 
small parameter, $\omega$ an eigenfrequency
and $(r,s)$ an eigenvector. Substituting it into 
eq.~(\ref{travNLS}) and keeping the linear terms in $r$ and $s$, one obtains the linear eigenvalue problem (EVP)
\begin{eqnarray}
\left[
\begin{array}{cc}
\mathcal{L} & u^2          \\
 -(u^{\ast})^2 & -\mathcal{L}^{\ast}
\end{array}
\right]
\left[
\begin{array}{cc}
r(x)\\
s(x)
\end{array}
\right]
=-\omega
\left[
\begin{array}{cc}
r(x)\\
s(x)
\end{array}
\right],
\label{EVP}
\end{eqnarray}
where $\mathcal{L}=iv\partial_x-\partial_{xx}-\mu+2|u|^2+\alpha\delta(x)$ and $\mathcal{L*}=-iv\partial_x-\partial_{xx}-\mu+2|u|^2+\alpha\delta(x)$. Note that eigenvalues in Hamiltonian systems come in complex quartets \cite{marsden}. In particular, in the realm
of the Schr{\"o}dinger systems (\ref{NLS}), if $\omega$ is an eigenfrequency of (\ref{EVP}) with a corresponding eigenfunction $[r(x)\, s(x)]^T$ and $T$ represents the transpose, then $(-\omega)$, $(-\omega^*)$ and $\omega^*$ are also eigenfrequencies with corresponding eigenfunctions $[s(-x)\, r(-x)]^T$,  $[s(x)^* \, r(x)^*]^T$ and $[r(-x)^* \, s(-x)^*]^T$, using the fact that $(u(x)^2)^*=u(-x)^2$ and $\mathcal{L}^*f(x)=\mathcal{L}f(-x)$. A solution $u(x)$ is therefore stable if and only if $\mathrm{Im}(\omega) = 0$ for all eigenfrequencies $\omega$.  

\begin{figure*}[tbhp]
\centering
\vspace{3cm}
\includegraphics[width=18cm,clip=]{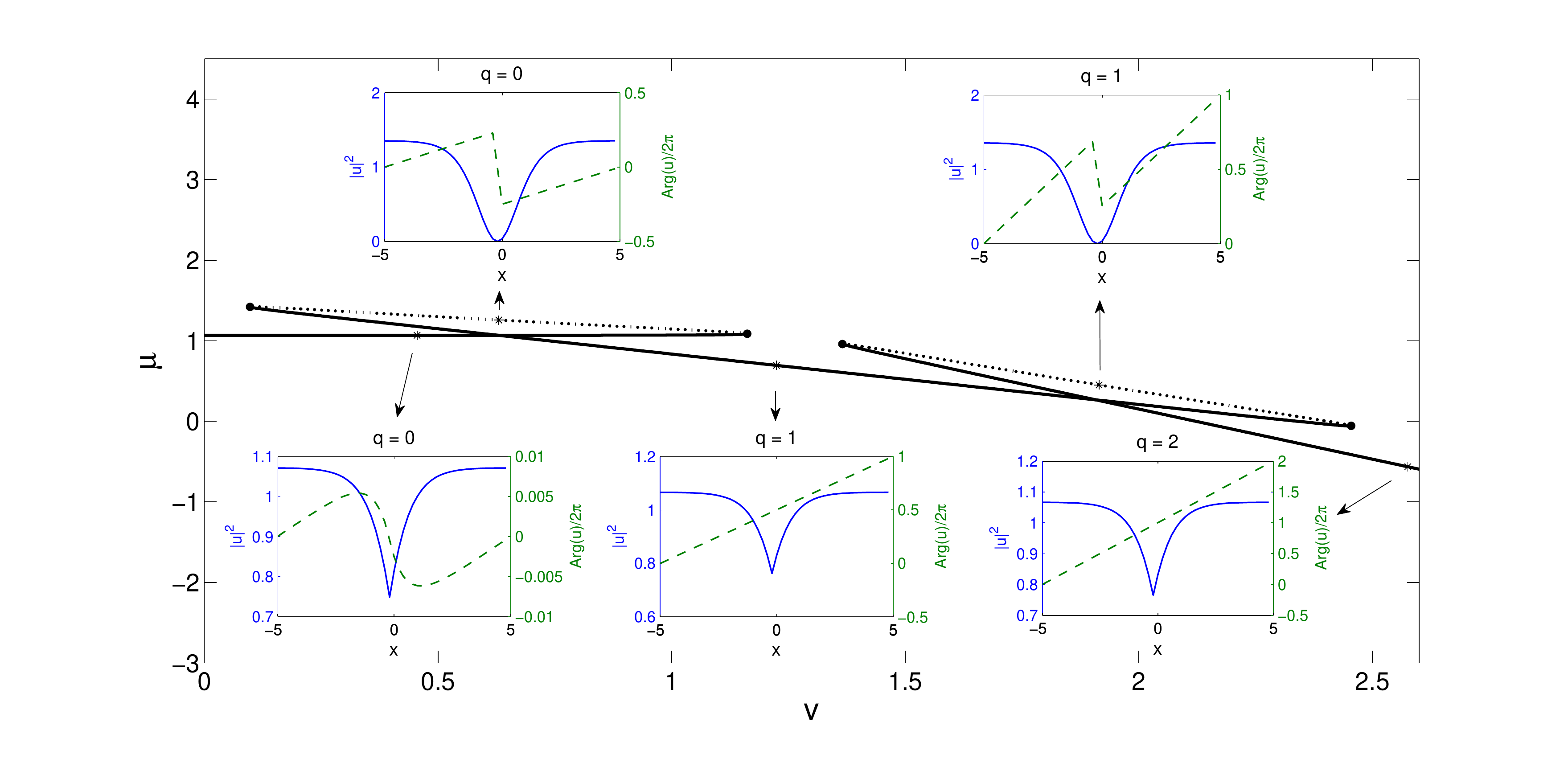}
\caption{Bifurcation diagram of the steady flow solution for a model system with $\alpha=0.5$ and $N_{L}=2L=10$. Bold solid and dotted lines correspond to
stable and unstable solutions, respectively. The insets show the time independent solution profiles in the ring trap along the branches for velocities corresponding to
the position of the crosses. Solid and dashed lines in the insets show the magnitude and phase of the solutions. For each profile the solution charge $q$ is given.}
\label{bifL5}
\end{figure*}

\section{Numerical Results and Connections to Theory}

In Figure~\ref{bifL5} we show one of the principal results of the present work, namely the bifurcation diagram of superfluid flow for varying velocity $v$ starting from the static solution $v=0$ of (\ref{NLS}) for a system with $N_L=2L=10$ and $\alpha=0.5$. 
As we fix the norm, the bifurcation diagram is depicted in $\mu$ as a function of $v$. From Figure~\ref{bifL5} we observe that the solution experiences many saddle-center bifurcations (turning points)
as indicated by black dots. Since steady flows do not exist beyond the turning points (i.e. for either larger or smaller $v$ for the respective turning point), the abscissa of the bifurcation points corresponds to critical velocities $v_c$. Note that hysteresis in one-dimensional rings and
in optical lattices has been reported before in \cite{muel02}, where it was argued that superfluidity  can be naturally viewed as a hysteretic response to rotation.

In a ring trap geometry the phase of the BEC circulates around the centre by an integer multiple of $2\pi$ (see figure~\ref{bifL5}). The so-called topological charge $q$ corresponds to how many times
the phase winds along the ring. Macroscopic states with different $q$ have distinct energies and the effect of $q$ on persistent flow has been recently studied experimentally \cite{moul12}.
Considering the phase of the solutions along the branch in figure~\ref{bifL5}, it is interesting to note that the topological charge jumps {along the branch segments
that correspond to decreasing velocity $v$.} More precisely, $q$ increases at the points where the density at the obstacle vanishes. Hence, the solutions for all values of $q$ are smoothly connected along the diagram. In Fig.\ \ref{bifL5}, the upper insets show the density and the phase profile right before the charge jump (phase slip). 

\begin{figure}[tbhp]
\centering
\includegraphics[width=10cm]{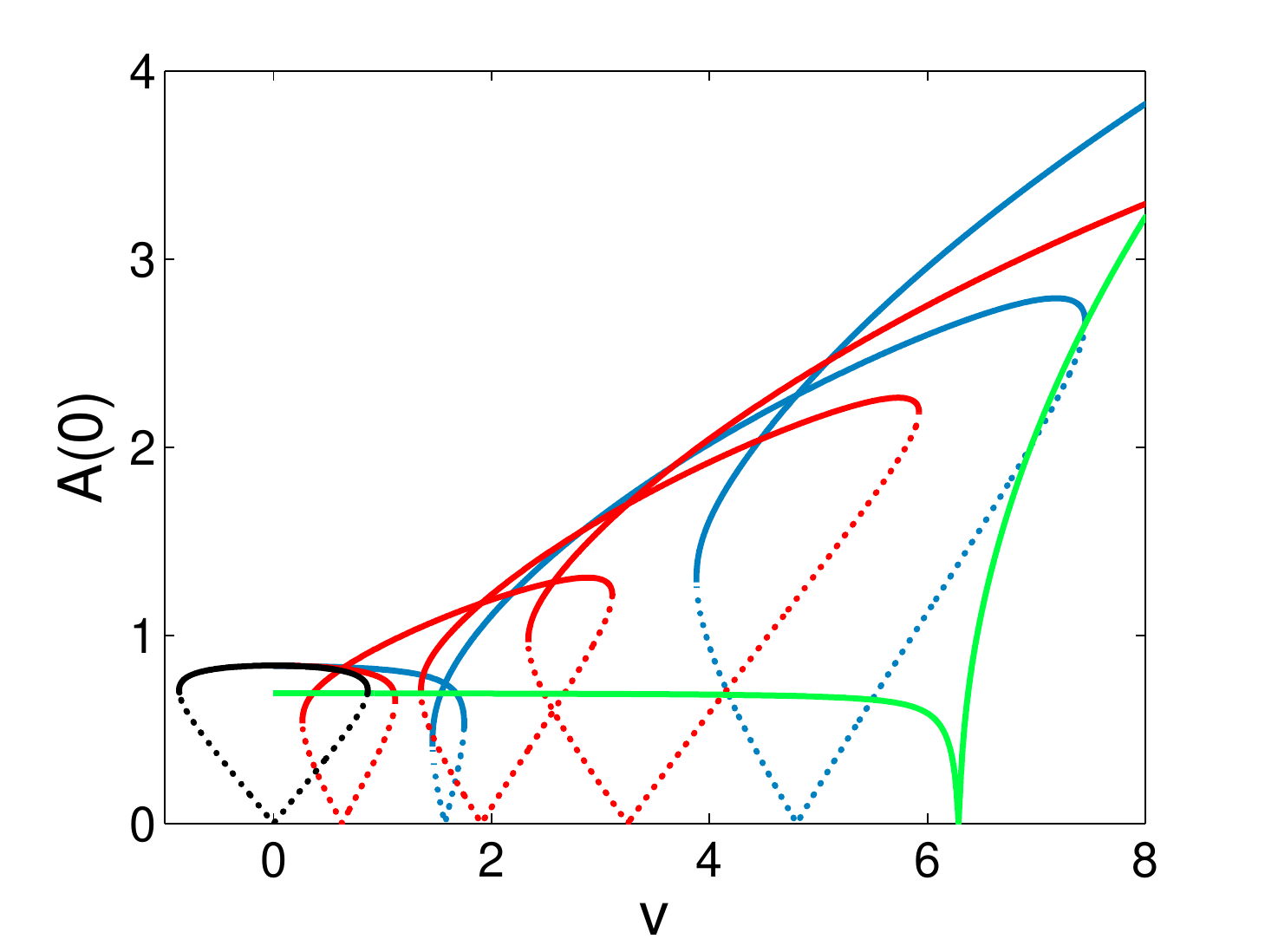}
\caption{(Colour online) Connection between ring systems with finite lengths $L$ (coloured) and the infinite domain (black line). We plot $A(0)=|u(0)|$ as a function of $v$ for $\alpha=0.5$ for several values of $L$, i.e.\ $\infty$ (black), $5$ (red), $2$ (blue), $1/2$ (green). Solid and dotted curves represent stable and unstable solutions, respectively. }
\label{comp}
\end{figure}

To provide a better understanding of the relation between the bifurcation diagram in ring systems (figure~\ref{bifL5}) and that of the infinite domain, which only has one saddle-center
bifurcation \cite{Hakim}, we now study bifurcation diagrams for different values of the domain length $L$. For this calculation it is preferable to fix the chemical potential $\mu$ and
let the solution norm vary. Using $\alpha=0.5$ and $\mu=1$, the results are shown in Fig.\ \ref{comp}. Plotted is the square-root density of the stationary solution $|u(0)|$ against the velocity.

It is known that in the infinite domain, the critical velocity corresponds to a saddle-center bifurcation between a dark soliton pinned to the obstacle and the uniform solution that is modified due to the inhomogeneity $U(x)$ \cite{Hakim}. The continuation diagram of the solution in this case forms a loop with its symmetric counterpart (note that Eq.\ (\ref{travNLS}) is invariant under the transformation $v\to-v$ and $x\to-x$) shown as black curve in Fig.\ \ref{comp}. As $v$ varies further, one will go around in the closed loop.

In a ring system, when $L$ is finite, we do not obtain a closed loop, but have connected `loops' instead. The situation, when the curves touch the horizontal axis for finite $L$, corresponds to the creation of dark-soliton-like states at the position of the impurity. Exactly at these points the topological charge increases. While in the infinite domain there is only one velocity point, where dark soliton-like-states can be created by the impurity, there are several of these points in the ring systems. The implications of this feature lead to
complex dynamics as will be discussed further below.

With decreasing length $L$ the initial `loops' become smaller and the distance between two consecutive `loops' increases. There will be critical values of $L$ when the shrinking `loops' become points, i.e.\ pairs of saddle-node bifurcations collide. In that case, when one decreases the length further, the corresponding `loops' will disappear as is the case in the green curve in figure~\ref{comp}. 

\begin{figure}[tbhp]
\centering
\includegraphics[width=8cm]{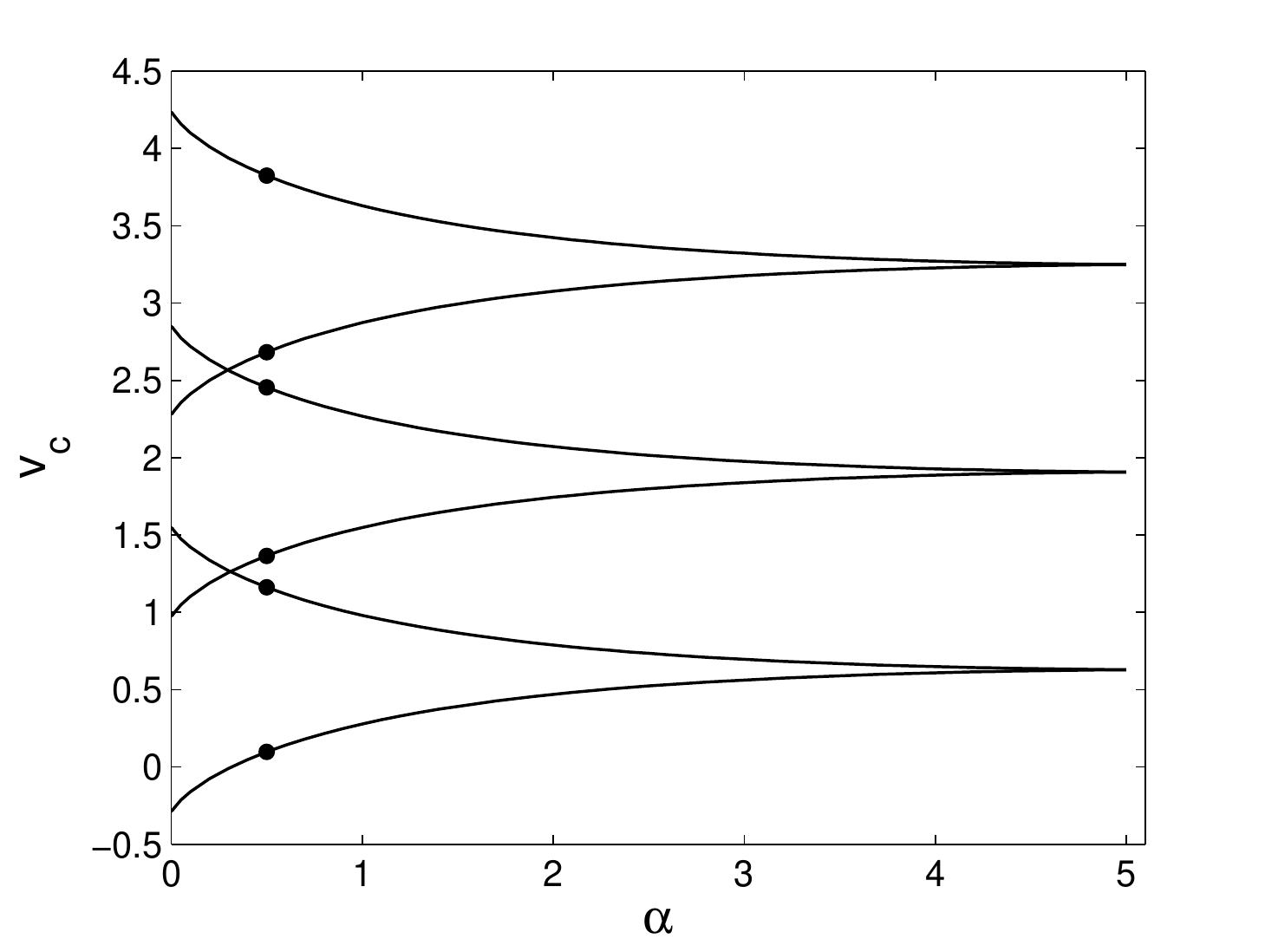}
\caption{The critical velocities $v_c$ as functions of the potential strength $\alpha$, in a ring with $L=5$. The black dots correspond to those depicted in figure~\ref{bifL5}. Above a certain critical strength the potential can move with any velocity without breaking the superfluidity.
This is because the critical points pairwise merge, as shown in the
figure. This is done 
in a way reminiscent of a swallowtail catastrophe 
surface~\cite{swallow}
in the context
of Fig.~\ref{bifL5} and results in a monotonic dependence
of the chemical potential $\mu$ vs.\ the velocity $v$ and the absence
of critical points.
}
\label{varied1}
\end{figure}

We have also studied the effect of varying the potential strength $\alpha$ on the existence of steady state solutions. From our computations shown in figure~\ref{varied1}, we
obtain that as $\alpha$ increases, the bifurcation `loops' are getting smaller and two consecutive saddle-center bifurcation points get closer to each other and then finally disappear at
some value $\alpha$. This implies that there is a critical potential strength parameter $\alpha_{cr}$ above which there is no saddle-node bifurcation, i.e.\ the obstacle can move with any velocity. 
This may be interpreted as a condition when the obstacle is strong enough to pin the ring BEC such that moving the obstacle means moving the BEC as a whole and hence there is no relative velocity between the two. This informs us that there is a limiting $\alpha$ above which it is unfavourable for the BEC to have a non-vanishing density at the position of the barrier, i.e.\ the BEC moves together with the barrier. In this case, the response of the BEC to a moving barrier should not be distinguishable from that of a thermal gas. 
The green curve ($L=1/2$) in figure \ref{comp} also corresponds to such a situation. This is a feature that is not present in the infinite domain. 

The bifurcation diagrams can be analysed as follows (a similar derivation was presented in the Supplemental Material of \cite{Cominotti14} without connection with bifurcations that occur in the system). Using the Madelung's transformation $u(x)=A(x)\mathrm{e}^{i\varphi(x)}$, one obtains from the static version of (\ref{travNLS})

\begin{eqnarray}
&A\varphi_{xx}=vA_x-2A_x\varphi_x,\label{eq1}\\
&A_{xx}=A\varphi_x^2-\mu A+A^3-vA\varphi_x+U(x)A. \label{eq2}
\end{eqnarray}
Multiplying (\ref{eq1}) with $A$ and integrating  yields
\begin{equation}
\varphi_x=\frac{v}2-\frac{C_1}{A^2},
\label{eq2t}
\end{equation}
where $C_1$ is a constant of integration, which can be directly taken to be any number in the infinite domain \cite{Hakim}. 
Notice that Eq.~(\ref{eq2t}) is suggestive (as discussed above in connection
with Fig.~\ref{comp}) of
the fact that where $A(0) \rightarrow 0$, sharp gradients in the phase
may arise; cf. the top profiles in Fig.~\ref{bifL5}.
For a $\delta$-potential $U(x)$, one then obtains from
(\ref{eq2}) the first integral

\begin{equation}
A_x^2=\frac1{4A^2} \left[ 2A^6-\left(4\mu +v^2\right)A^4+4C_2A^2-4C_1^2 \right],
\label{eq3}
\end{equation}
with $C_2$ being a constant of integration, and boundary conditions
\begin{equation}
A(0^+)=A(0^-),\,A_x(0^+)-A_x(0^-)=\alpha A(0).
\label{b2}
\end{equation}
Due to the symmetry ($x\to-x$), the latter equation is equivalent to $A_x(0^+)=\alpha A(0)/2$.
Using the equations in (\ref{b1}), which are equivalent to $A(-L)=A(L),\,A_x(\pm L)=0,$ and (\ref{b2}), it is straightforward to obtain from (\ref{eq3}) evaluated at $x=0,L$ that
\begin{eqnarray}
K_1 & = A ( 0 ) 
 ^{2}\left(2A(0)^{2}-\left({\alpha}^{2}+{v}^{2}+4\mu\right)\right),\nonumber\\
 K_2 &= A ( L
 )^{2}\left(2A ( L )  ^{2}-\left({v}^{2}+4\mu\right)\right),\nonumber\\
K_3&=A ( 0 )  ^{4}\left(2A(0)^{2}-\left({\alpha}^{2}+{v}^{2}\right)\right),\,
K_4=A(L)^{4}\left(2A(L)^{2}-\left({v}^{2}+4\mu\right)\right),\nonumber\\
C_1^2&=
{\frac { A ( L )  ^{2} A
 ( 0 )^{2} \left(K_1 -K_2 \right) }{ 4\left( A ( L)^{2}- A (0) ^{2}\right)}},\, C_2=
{\frac {K_3-K_4-4A(0)\mu}{ 4\left( A ( L)  ^{2}- A
 ( 0 )^2  \right) }}.
\end{eqnarray}

Let
\begin{equation}
Y(y)=\int \frac{dy}{\sqrt{2y^3-\left(4\mu +v^2\right)y^2+4C_2y-4C_1^2}},
\end{equation}
which can be expressed in terms of the incomplete elliptic integral of the first kind \cite{olve10}. Then, the solution of (\ref{eq3}) is
\begin{equation}
A(x)=\sqrt{Y^{-1}\left(x+Y(A(0)^2)\right)},\,x>0,
\label{eq4}
\end{equation}
and $A(-x)=A(x)$. Hence, $A(L)$ can be written in terms of $A(0)$. Finally, using (\ref{eq2t}), one obtains the nonlinear algebraic equation that will yield the diagrams
in Fig.\ \ref{comp}, i.e.\
\begin{equation}
\frac{vL}{2}-q\pi=\int_0^L{\frac{C_1}{Y^{-1}\left(x+Y(A(0)^2)\right)}\,dx},
\label{eqfinal}
\end{equation}
where $q\in\mathbb{Z}$ is the topological charge. Figure~\ref{bifL5} can be obtained similarly from solving (\ref{eqfinal}) simultaneously with the constraint
$\int_0^L{{Y^{-1}\left(x+Y(A(0)^2)\right)}\,dx}=N_{L}/2$ for $\mu$ and $A(0)$.

\begin{figure}[tbhp]
\centering
\subfigure[$\,q=0$]{\includegraphics[width=7.7cm]{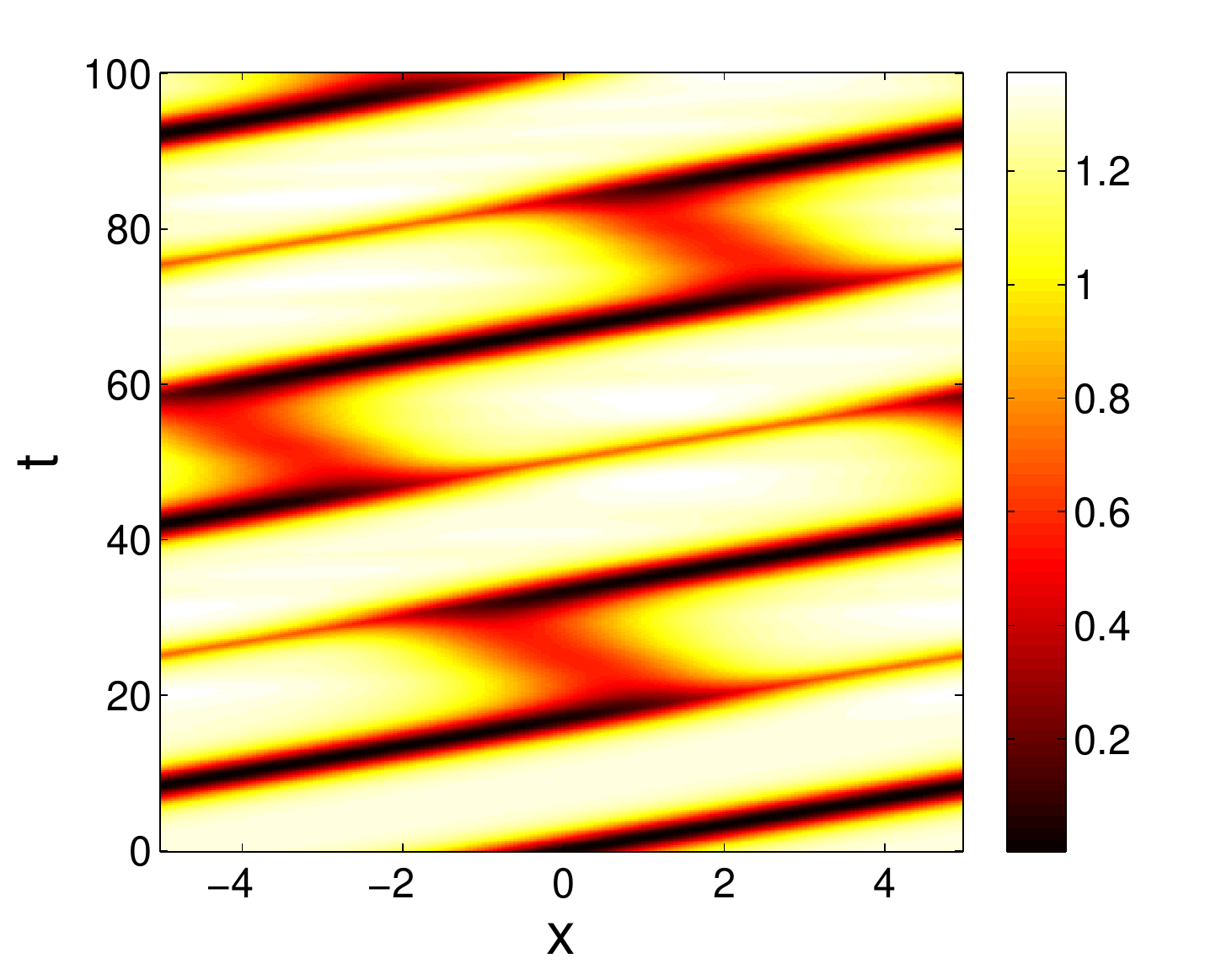}}
\subfigure[$\,q=1$]{\includegraphics[width=7.7cm]{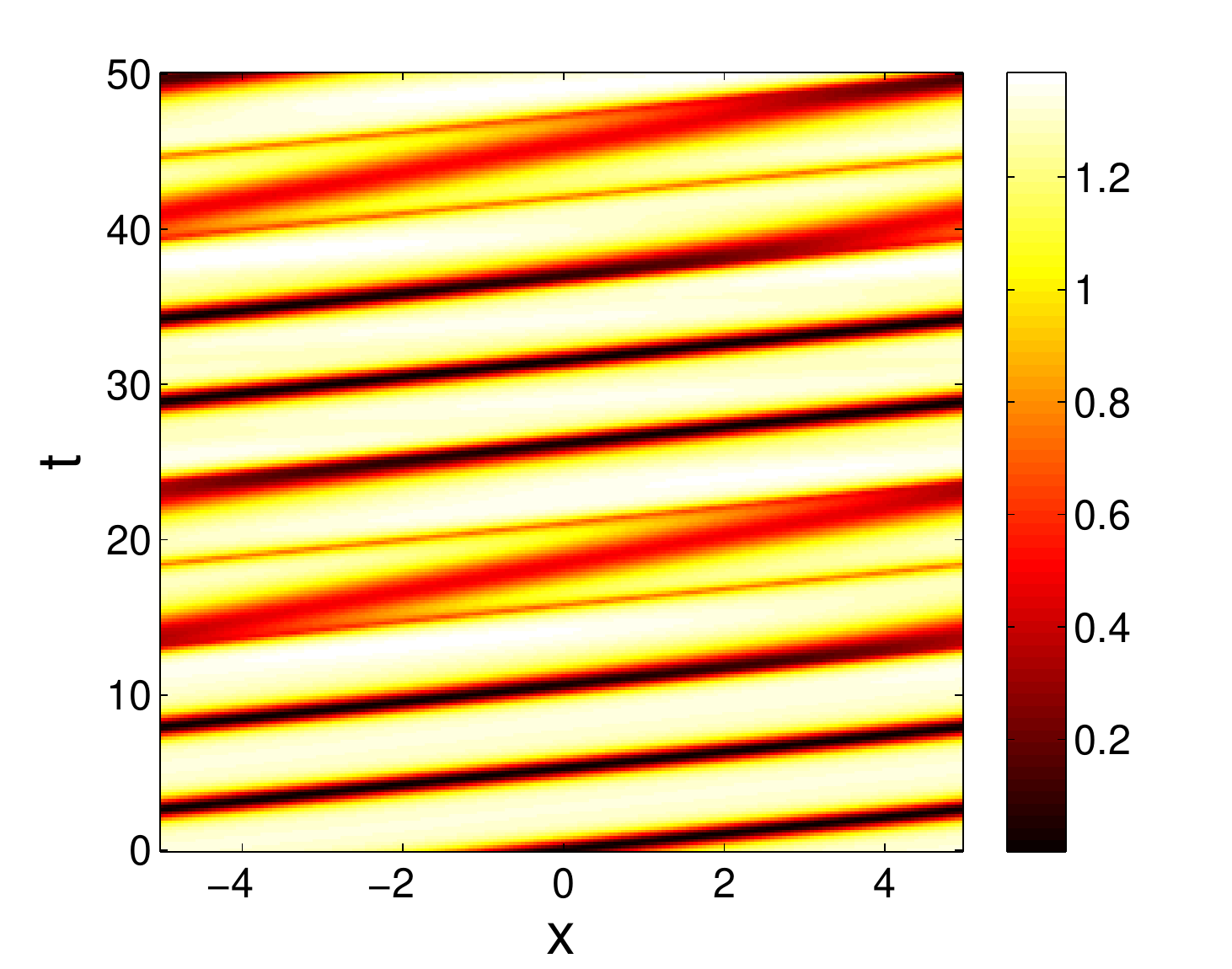}}
\caption{(Colour online) Time dynamics of the two unstable solutions shown in the insets of figure~\ref{bifL5} for $q=0$ and $v=0.6$ (a) and $q=1$ and $v=1.9$ (b). In both panels,
a soliton is created and moves around the ring with constant velocity, while the density of the BEC shows a periodic pattern. Depicted is the density $|u(x,t)|^2$.}\label{evoluns}
\end{figure}

\begin{figure}[tbhp]
\centering
\subfigure[]{\includegraphics[width=7.7cm]{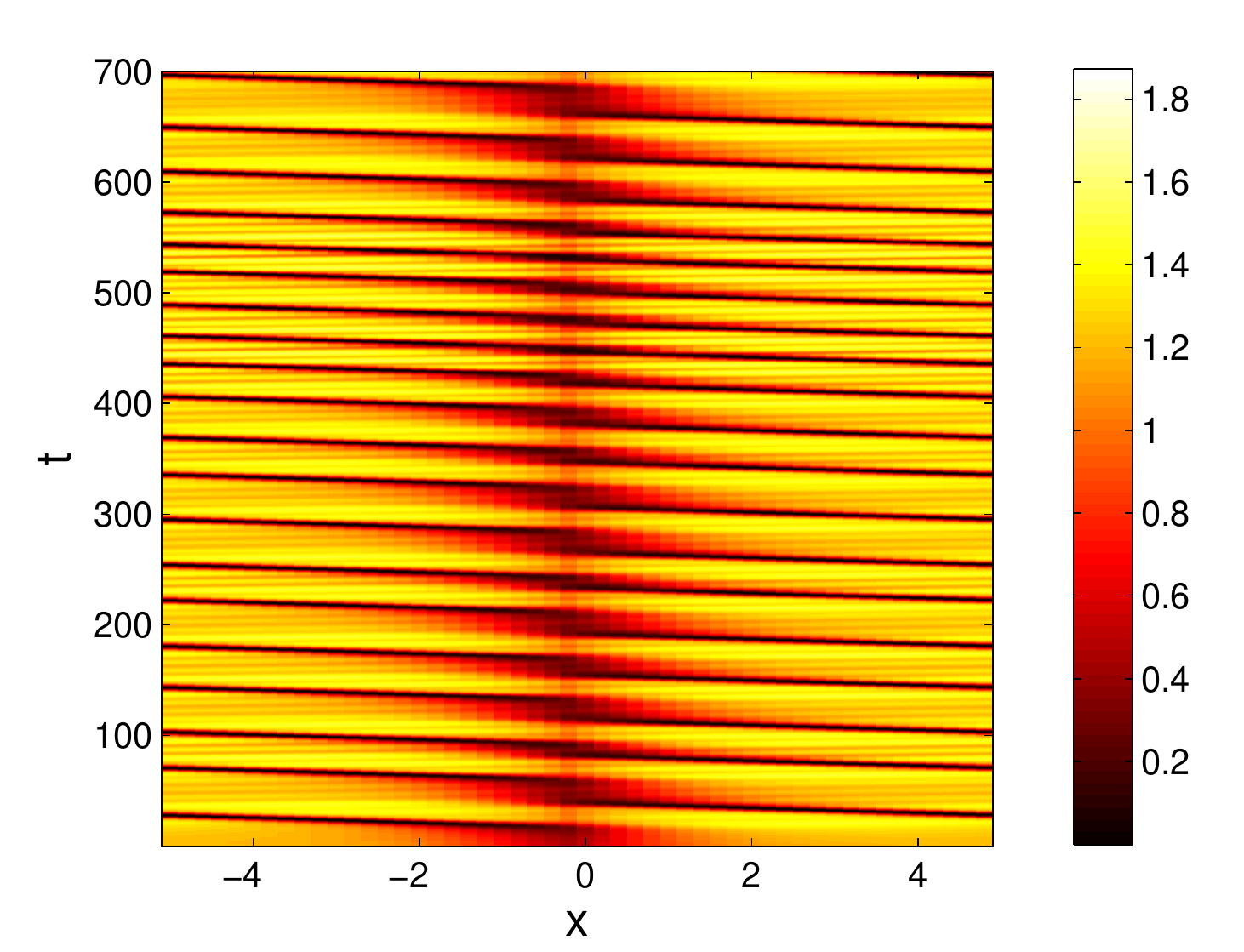}}
\subfigure[]{\includegraphics[width=7.7cm]{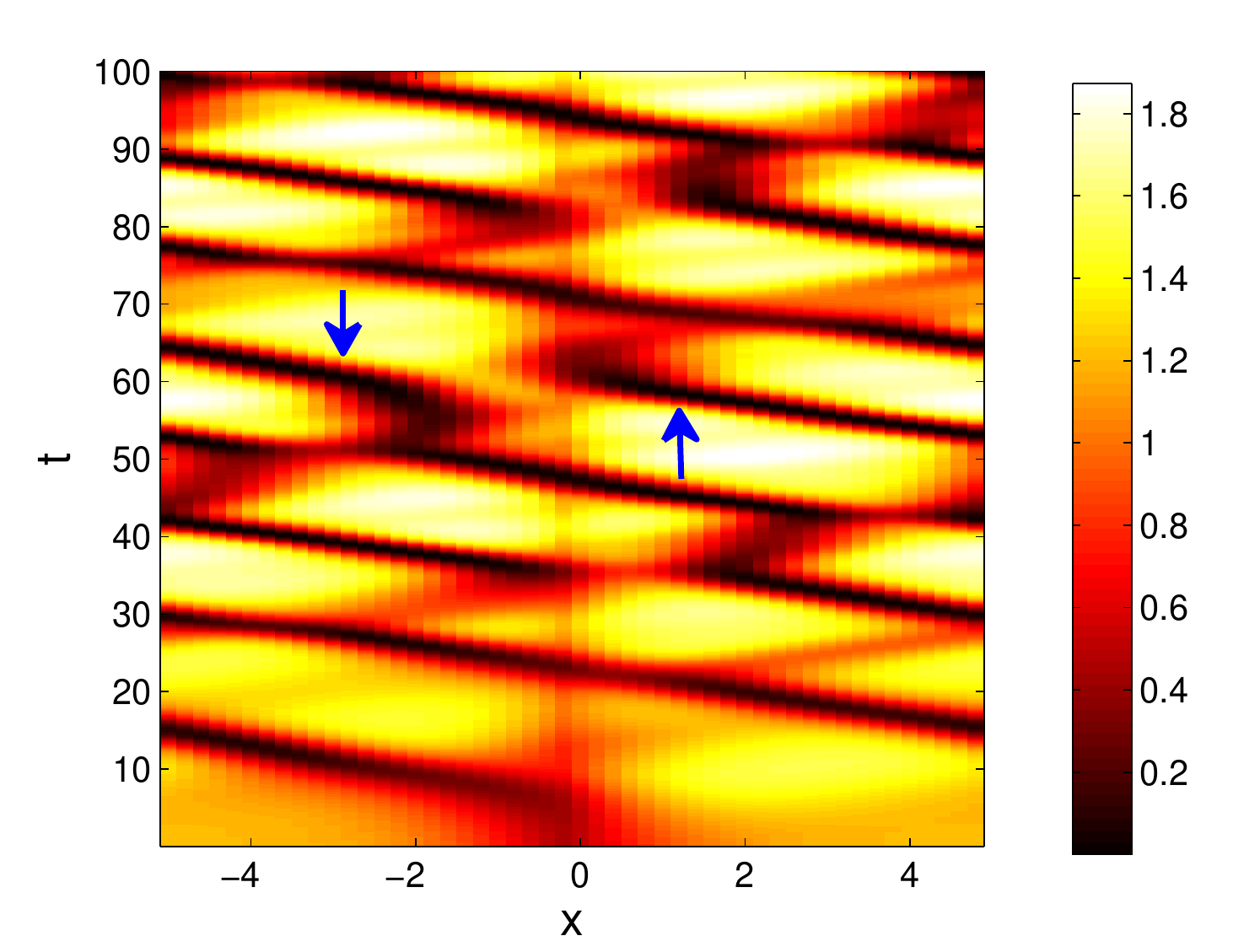}}
\caption{(Colour online) Numerical integration of eq.~(\ref{travNLS}) at $v=1.2+\Delta v$ close to the bifurcation point. (a) For $\Delta v=0.02$ a soliton is created at
the position of the impurity, after one round-trip the soliton is trapped by the impurity and released after some time. (b) For $\Delta v=0.6$ there are one or two solitons
within the trap. Arrows in the figure indicate the presence of two solitons at one instant of time. See text for details.}\label{evolL5}
\end{figure}

It is then interesting to investigate the dynamics of solutions of eq.~(\ref{travNLS}) in two specific regions of the bifurcation diagram. To be more precise, we study the time-evolution of unstable solutions under small perturbations (Figure~\ref{evoluns}) and discuss the time dynamics of solutions for velocities beyond the critical values (Figure~\ref{evolL5}), i.e.\ below or above which the corresponding steady state solutions do not exist. 


For the two unstable solutions depicted in the insets of Figure~\ref{bifL5}, we show the time evolution dynamics in laboratory frame in Figure~\ref{evoluns} by plotting the density distribution along the ring. 
In both situations a dark soliton is released from the inhomogeneity after initially travelling along the obstacle. 
The detachment of dark soliton from the 'pinning' potential is the only dynamics of instability that we observed in all our simulations. Whether the soliton travels ahead or behind the obstacle depends sensitively on the perturbation. Note as well that the density of the cloud shows that after detachment the dark soliton in panel (b) moves faster than that in panel (a). 

We then analyse the dynamics beyond a critical value, i.e. we take the steady state solution at a critical velocity $v_c$ as the initial condition and then compute the evolution for
$v=v_c+\Delta v$.

In figure~\ref{evolL5}(a) we show the dynamics in moving coordinate frame for clarity with $v=v_c+0.02$ for the second bifurcation point in Figure~\ref{bifL5} at $v_c=1.2$. We observe that a dark soliton is
emitted from the impurity along the evolution. In striking difference to the infinite domain \cite{Hakim}, the released dark soliton interacts periodically with the impurity,
which then traps the soliton for some time (e.g.\ between $t=40$ and $t=50$) before releasing it again. Note that the trapping time after each round trip is not constant.
For a wide interval of $\Delta v>0$ ($1.2 < v < 1.3$)
we obtain similar dynamics with only one soliton present in the annular BEC. In the ``laboratory frame'' (where the obstacle is rotating), this situation implies that the emitted dark soliton will tend to be standing, even though it 
also slowly drifts due to the temporary entrapment that it suffers from 
the moving obstacle.

When $\Delta v=0.6$, one or two dark solitons can be created within the ring, 
as is shown in figure~\ref{evolL5}(b). For the first few time units, there is only one soliton
present within the ring. After some time an additional dark soliton can be created, and as a result at the time instant indicated with arrows,  two solitons exist in the trap. The presence of several solitons in the trap and repeated interaction with the impurity can lead to complex dynamics including collisions between the solitons or annihilation of the soliton by the impurity.

\section{Discussion and Future Challenges}

In current state of the art ring traps that have been created e.g. by a spatial light modulator (SLM) \cite{moul12} or by magnetic potentials \cite{prit12}, 
the impurity can easily be added 
through a focused blue detuned laser beam or by the SLM as well. In the setup of \cite{rama11,ryu13} the impurity is present and excitations have indeed been observed \cite{wright13}. 
The predictions of this paper are therefore directly relevant for the effects observed in current experiments.

In summary, we have studied the creation of dark solitary waves
(and more generally the generation of phase slips/persistent currents) 
in a system with periodic boundary conditions. In contrast to the infinite 
domain case, for our bounded domain setting we find the existence of several
critical velocities corresponding to different 
charges $q$ of the stable solution. 
A somewhat unexpected feature was also the existence of sufficiently
narrow domains or sufficiently strong obstacles for which no critical
velocity could be identified. 
The ability to create coherent structures by increasing the velocity or to annihilate them through the impurity allows the creation --via the bifurcation
diagram presented herein-- of a controllable number of 
$2 \pi$ phase windings within the ring trap. 
The analytical tractability of this formation through the quasi-1D
theoretical formulation proposed herein is a feature adding to the
controllability of the process.
The exact dynamics of the resulting structures 
can be highly complex including possible collisions and interactions and 
will be an interesting object for further study. 
It is especially relevant to systematically
extend such considerations to higher dimensional contexts not only
in 2D but also in 3D. In higher dimensions, the situations are distinct as superfluid does not necessarily rotate like a solid body, see e.g.,\ \cite{piaz09} for numerical studies of flow dissipation in 2D in the mean field regime in the presence of a static barrier, \cite{yaki15} for superfluid with $q-$topological charge in the presence of rotating barrier and \cite{muno15} for persistent currents in the 3D case.

HS, MS, APM and LH acknowledge the University of Nottingham for financial support through an EPSRC Bridging the Gaps grant. APM acknowledges CONACYT, this project was supported by EPSRC grant EP/K023624/1 and by the European Comission grant QuILMI - Quantum Integrated Light Matter Interface (No 295293). PK and IL also acknowledge EPSRC (EP/I017828/1) and EU (FET Proactive grant 601180 “Matter- Wave”). P.G.K. gratefully acknowledges the
support of NSF-DMS-1312856, as well as from the BSF under grant
2010239, and the ERC under
FP7, Marie Curie Actions, People, International Research
Staff Exchange Scheme (IRSES-605096).

\end{document}